\documentclass[journal]{IEEEtran}
\usepackage{color}
\usepackage{amsmath,amssymb,amsfonts}
\usepackage{algorithm}
\usepackage{algorithmic}
\usepackage{tabularx}
\usepackage{graphicx}
\usepackage[T1]{fontenc}
\usepackage{array}
\usepackage{stfloats}
\usepackage{float}
\usepackage{subfigure}
\usepackage{nccmath}
\usepackage{setspace}
\usepackage{booktabs}
\usepackage{mwe}
\usepackage{ulem}
\usepackage{cite}
\usepackage{amsfonts}
\usepackage{kotex}
\usepackage{pifont}
\newcommand{\xmark}{\ding{55}}
\usepackage{textcomp}
\usepackage{xcolor}
\usepackage{balance}
\newtheorem{theorem}{Theorem}
\newtheorem{defi}{Definition}
\usepackage{xspace}

\newcommand{\BfPara}[1]{{\noindent\bf#1.}\xspace}

\def\BibTeX{{\rm B\kern-.05em{\sc i\kern-.025em b}\kern-.08em
    T\kern-.1667em\lower.7ex\hbox{E}\kern-.125emX}}

\begin{document}
\bibliographystyle{jcn}

\title{Neural Myerson Auction for Truthful and Energy-Efficient Autonomous Aerial Data Delivery}

\author{Haemin Lee, Sean (Seok-Chul) Kwon, Soyi Jung, and Joongheon Kim
	\thanks{This research is supported by National Research Foundation of Korea (2019R1A2C4070663, 2021R1A4A1030775).}
    \thanks{Haemin Lee and Joongheon Kim with the School of Electrical Engineering, Korea University, Seoul 02841, Republic of Korea; e-mails: haemin2@korea.ac.kr, joongheon@korea.ac.kr.}
	\thanks{Soyi Jung is with the School of Software, Hallym University, Chuncheon, Korea e-mail: sjung@hallym.ac.kr.}
	\thanks{Sean Kwon is with the Department of Electrical Engineering, California State University, Long Beach, CA, USA e-mail: sean.kwon@csulb.edu.}
    \thanks{S. Jung and J. Kim are the corresponding authors of this paper.}%
}

\markboth{JOURNAL OF COMMUNICATIONS AND NETWORKS}
{Lee \lowercase{\textit{et al}}.: Neural Architectural Myerson Auction for Truthful Aerial Delivery Mobility} 
\maketitle

\begin{abstract}
A successful deployment of drones provides an ideal solution for surveillance systems. Using drones for surveillance can provide access to areas that may be difficult or impossible to reach by humans or in-land vehicles gathering images or video recordings of a specific target in their coverage. Therefore, we introduces a data delivery drone to transfer collected surveillance data in harsh communication conditions. This paper proposes a Myerson auction-based asynchronous data delivery in an aerial distributed data platform in surveillance systems taking battery limitation and long flight constraints into account. In this paper, multiple delivery drones compete to offer data transfer to a single fixed-location surveillance drone. Our proposed Myerson auction based algorithm, which uses the truthful second-price auction (SPA) as a baseline, is to maximize the seller’s revenue while meeting several desirable properties, i.e., individual rationality and incentive compatibility while pursuing truthful operations. On top of this SPA-based operations, a deep learning based framework is additionally designed for delivery performance improvements.
\end{abstract}



\section{Introduction}\label{sec:1}
\begin{figure}[!]\centering
    \includegraphics[width=1\columnwidth]{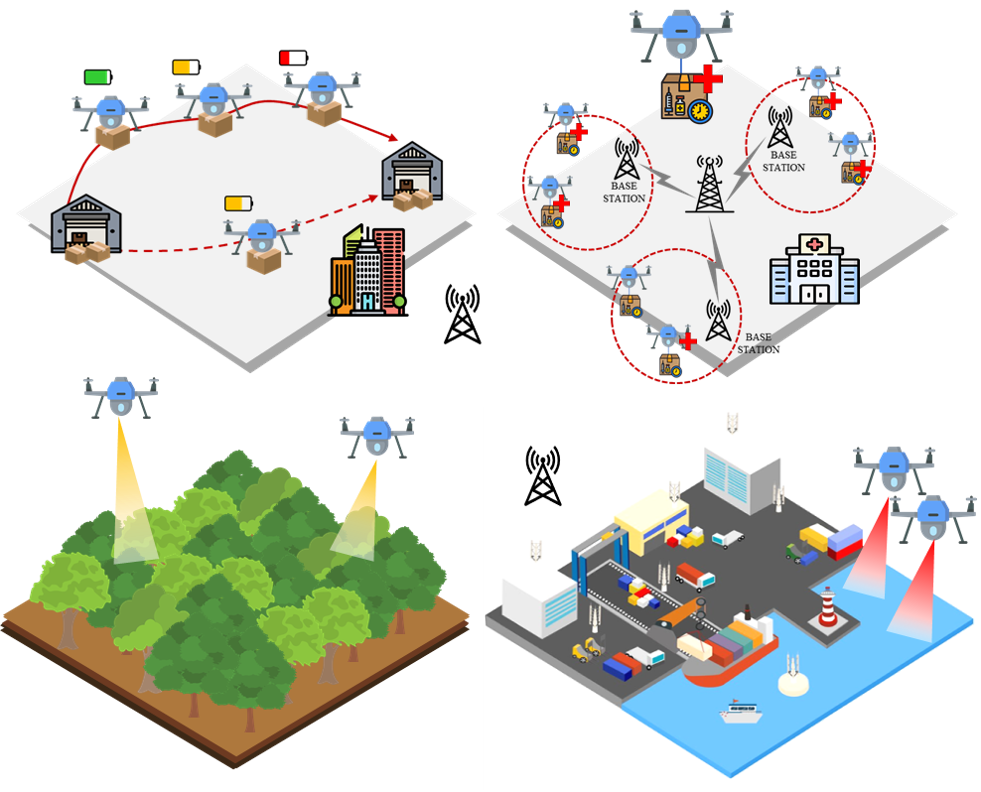}
    \caption{Drone for delivery usages.}
    \label{fig:usage}
\end{figure}

In recent years, an increasing number of enterprises, including Amazon, DHL, and Federal Express (FedEx), has been testing the viability of incorporating drone delivery into their commercial package delivery services~\cite{zhang2017constrained}. Due to the natural trait of drones which can swiftly move and optimize their path to quickly complete their mission, their role is getting more extensive and diverse in delivery sector.
Fig.~\ref{fig:usage} shows the drone delivery usages from simple warehouse logistic to various applications. Especially in the ongoing pandemic situation, drones are used for timely vaccine distribution during novel coronavirus (COVID-19) and future pandemics~\cite{verma2021vacochain}. Furthermore, drones also facilitate the inspection of hard-to-access areas configuring situational awareness~\cite{access19geraldes, jung2021adaptive, kim2021stabilized}, which is important for many applications such as remote sensing, search and rescue, and disaster response. As presented in Fig.~\ref{fig:usage}, monitoring and data collection of smart sensor in mountainous areas or in harbor facilities are also possible.

This paper proposes a data delivery service by the drone network in a surveillance system. With advances to AI, modeling, and simulation technologies, it has become essential to analyze data to extract potential values and create additional services~\cite{na2017energy, park2021joint}. In this situation, a large amount of data accumulation is fundamentally required, including various range of information. In a series of processes in a data platform, i.e., data collection, data processing, and data analysis., drones can collect raw data and deliver it to the target point operating as a component of the platform. The monitoring data enables facility infrastructure management, spatial information acquisition, and adequate response to rising problems. In this paper, a surveillance drone is presented, locating at specific region, to monitor important points of interest (PoI). We assume the situation where the existing communication network is temporarily destroyed or overloaded due to unexpected events like natural disasters. It is sometimes unrealistic to allow surveillance drone to transmit or relay monitoring data to nearby base station or other surveillance drone. In order to achieve the purpose of surveillance, the collected data has requirement that the total time should be no greater than a given maximum delay time to ensure the freshness of the data. Therefore, the accumulated data in destroyed network infrastructure should be delivered to areas where cellular communication is available.
In this situation, transferring data directly from hovering surveillance drones to base stations is not efficient due to low power-to-data transmission efficiency. If the distance between the surveillance spot and the final destination, i.e., the base station where the data should be transferred is far, path loss is severe proportionally affecting reception sensitivity. 
In this regard, reliable and calamity-resilient communication infrastructure is needed to deploy drone application services effectively in poor communication conditions. 

This paper proposes an effective aerial data delivery surveillance platform with the delivery drone that directly transmits data by moving back and forth in the middle of two points. It might cause some extent of delay but more efficient in terms of the transmission rate. Surveillance drone collects data from its specific monitoring spots. Delivery drones usually hover around, and when they are eligible to transfer data,  they fly straight to the destination. Then, they return with the data to the nearby base station contributing as a data provider in the big data platform. As a sequence, flying drones can facilitate rapid information dissemination by broadcasting common files among ground devices despite of the communication adversity. For example, it can bring the public interest by spreading the necessary information to be aware in public. We assume that the third-party operator pays the commercial delivery drones in return for delivery. For this reason, the delivery drones are willing to use up their energy to transfer the data trying their best to fulfill the given task. However, due to the destroyed infrastructure scenarios, surveillance drone can not assign data to multiple delivery drones. In other words, the delivery drones compete to deliver the collected data to the destination. By taking an econometric approach, matching between the two drones can be formulate as an auction where delivery drones act as buyer and surveillance drone acts as seller and auctioneer. Throughout the process, the objective is to maximize the seller’s revenue and buyer’s utility in the same time. Then the surveillance drone hand over the collected data in its queue storage and gains the spatial benefit to take the new sensing data.

This paper operates the series of auction process with deep learning auction networks. As the Myerson auction achieves revenue-optimal by transforming the bids through monotonic transformation functions, our neural network borrows the concept. The amount of delivery data and final payment of the winner are calculated through the neural network with the collected bids. Based on our method, performance evaluation confirms that the auction based matching between the delivery drone and surveillance drone maximizes the sellers revenue compared to the SPA (Second-Price Auction).

\BfPara{Contributions}
The main contributions of this research are three-fold and are summarized as follows.
\begin{itemize}
    \item This paper proposes a novel drone deployment for efficient surveillance data delivery in harsh condition, which is the first attempt to the best of our knowledge.
    \item In addition, our algorithm is designed and implemented fundamentally based on SPA which is mainly used for truthful resource allocation.
    \item Moreover, our proposed SPA-based algorithm is improved using deep learning framework in order to maximize the seller's revenue and buyer's utility in the SPA-based truthful auction settings. Finally, the performance of deep learning-based auction structure is compared with the traditional SPA.
\end{itemize}


 
\section{Preliminaries}\label{sec:related}

\subsection{Auction-based Resource Allocation}\label{sec:3}
A traditionally well-known first-price auction (FPA) is a common type of auction that the bidder who submits the highest bid value to auctioneer (seller) is awarded and pays its bid value to the auctioneer.
Here, suppose that $N$ bidders, i.e., $b_{1},\cdots, b_{N}$, and $1$ auctioneer exist in the system, where the bid values are $v_{1},\cdots, v_{N}$. The auctioneer selects one bid value $v^{*}$ with $v^{*}=\max\{v_{1},\cdots, v_{N}\}$; and the winner bidder $b^{*}$ will be the bidder who submitted bid value $v^{*}$. Suppose that the second highest bid value is $v^{\dag}$. Then, the winner bidder $b^{*}$ does not need to pay $v^{*}$ amounts of bid values because slightly higher bid value than $v^{\dag}$ will guarantee the winning. Therefore, individual bidders need to be strategic in FPA. 
On the other hand, the other type of auction mechanism is called second price auction (SPA). With SPA, the mechanism for selecting a winner is equivalent to FPA, where the payment by the winner is not the winner's highest bid value but the second highest bid value. In the literature, the SPA is well-known for its 
turthfulness~\cite{tvt201905shin,infocom2017}. Therefore, SPA is widely used for truthful resource allocation in various distributed computing applications~\cite{srcho1,srcho2}. However, one drawback in SPA is that the SPA cannot achieve revenue-optimal, i.e., the auctioneer cannot obtain maximum benefits be because the second bid value will be given to the auctioneer rather than the highest.

In order to pursue revenue-optimal in SPA, various approaches have been studied. Among them, Myerson auction with the concept of virtual valuation is one of the well-known approaches~\cite{tvt201905shin, auction_tii}. In order to numerically formulate the virtual valuation, monotonic increasing functions are generally used. According to the advances in deep neural network (DNN) research, the Myerson auction computation procedure can be approximated with the form of DNN.
Therefore, this paper designs our proposed DNN-based autonomous aerial delivery scheduling algorithm with the name of neural-architectural Myerson auction.

\subsection{Related and Previous Work}
\subsubsection{Data Delivery Research in Multi-Drone Networks} 
There have been several research on data acquisition frameworks in wireless sensor networks using drones with the goal of increasing the efficiency of the data gathering efforts.
The proposed algorithm in \cite{say2016priority} introduces a priority-based frame selection scheme to suppress the number of redundant data transmissions between sensor nodes and the drones.
In addition, the algorithm in \cite{zhan2017energy, data_tii} utilizes the drones as mobile data collectors for the randomly deployed sensor nodes. In this research, the proposed algorithm jointly optimizes the wake-up schedules sensor nodes and the trajectories of drones in order to minimize the maximum energy consumption (i.e., min-max criteria for fairness).
Moreover, the proposed algorithm in \cite{gong2018flight} minimizes the drone's total flight time while allowing each sensor to successfully upload a certain amount of data.
Furthermore, the proposed algorithm in~\cite{singhal2021aerial} provides adaptive surveillance and event-telecast video streaming services from drones to ground control stations with WiFi-direct link scheduling its associated dynamic configuration settings.

Our aerial data delivery scheduling with neural Myerson auction computation in this paper differs in scenario that the delivery drones compete to directly transfer the data. This approach superior from the other approaches because Given that the above researches focus on priority and energy efficiency aspects of the drone-based data collection process, our methods works in extremely poor conditions by sustainably enabling data transmission. Furthermore, our deep learning-based auction reduces the costs through one optimal delivery drone selection in the data collection process. This paper is in line with leading research on drone-based data delivery networks in that it considers the energy and coverage of drones as the bid values of individual drones. However, it differs from others in that it enables data delivery even when the communication infrastructure is poor or destroyed.


\subsubsection{Previous Work in Auction-based Resource Allocation}\label{sec:application}
The auction approach is an useful intuitive method for solving distributed scheduling and resource allocation problems in a distributed and truthful way. There exists inherent uncertainty regarding valuations for both auctioneers/sellers and buyers/bidders.
The auctioneers is unsure about the values that bidders attach to the object being sold, i.e., the maximum amount each bidder is willing to pay. If the auctioneer knew the values precisely, it could just offer the object to the bidder with the highest value at or just below what this bidder is willing to pay. No bidder knows with certainty the values attached by other bidders and the knowledge of other bidders' values would not affect how much the object is worth to a particular bidder~\cite{krishna2009auction}.
With a massive volume of economic transactions is conducted through auctions, numerous research on limited resource allocation and scheduling problems has been conducted through auction-based computation processes~\cite{klemperer1999auction}.
In~\cite{dai2014price}, a carrier collaboration problem with pick and delivery requests is considered in order to reduce their transportation costs and consequently increase their profits. A multi-round pricing-setting based combinatorial auction approach is proposed to solve the problem. The proposed algorithm in~\cite{marinescu2013auction} sketches a self-organizing architecture for very large compute clouds and provides a relatively simple, scalable, and tractable solution to cloud resource allocation through the combinatorial auction. On the other hand, the proposed algorithm in~\cite{coltin2013online} introduces an auction-based scheduling algorithm that plans to transfer items between robots to conduct deliveries in a more efficient way. The algorithm runs online and replans in response to new requests, dead vehicles, and shared information. Authors in~\cite{truthful_tii} propose an auction-based incentive mechanism that achieves near-optimal long-term social welfare in collaborative computation offloading.

Among various auction-based scheduling and resource allocation algorithms, our considering Myerson auction is one of the most efficient revenue-optimal single-item auctions~\cite{myerson1981optimal}.
In order to numerically approximate the Myerson auction, DNN-based architecture can be utilized; and thus, learning-based Myerson auction algorithms for charging scheduling in wireless power transfer (WPT)-based multi-drone networks and electric vehicles are proposed in~\cite{tvt201905shin} and \cite{lee2021truthful}, respectively. 
In addition, the proposed algorithms in~\cite{zhu2020revenue} and~\cite{luong2018optimal} solve resource allocation problems using DNN-based auctions in mobile edge computing and wireless virtualization, respectively. 
Furthermore, the proposed algorithm in~\cite{kuo2020proportionnet} is for approximating auctions using deep learning to address the concerns of fairness while maintaining high revenue and strong incentive guarantees.

\section{System Model -- Autonomous Aerial Mobility}\label{sec:4}
\subsection{Overall Architecture} 
\begin{figure}[t!]
\centering
\includegraphics[width =1\columnwidth]{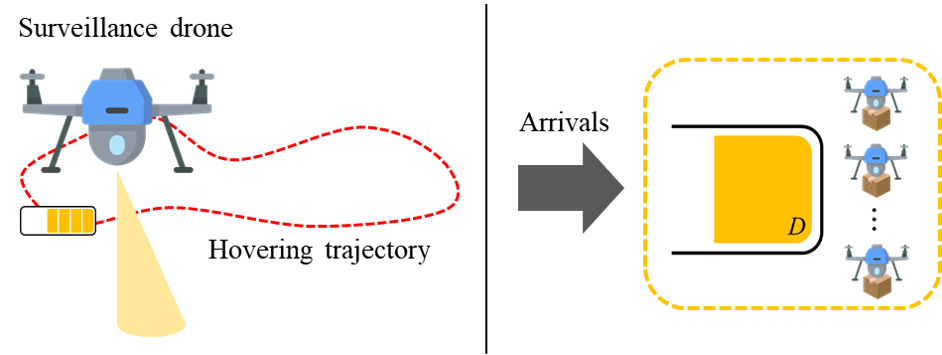} \\
 \centering (a) The corresponding queue model of SD.\\
 \centering \includegraphics[width =1\columnwidth]{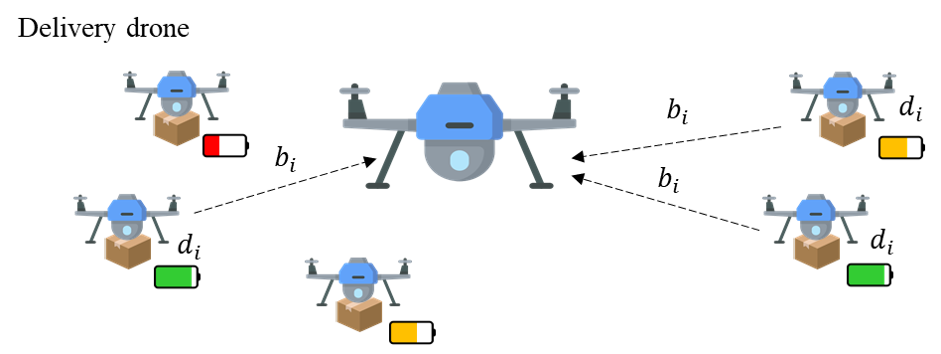}\\ 
(b) The bidding behavior of DD.\\
\caption{Drone operation model.}\label{fig:architecture}
\end{figure}

Our aerial data delivery for surveillance system consists of three elements, i.e., surveillance drone, delivery drone and base station. The computation during the auction is done on-device within a surveillance drone. Surveillance Drone $S$ collects monitoring image while hovering over a specific region and these surveillance drone is managed by third party operators. Surveillance drone hovers around its specific area and collects information in different observation angles and heights, not being affected by its surrounded obstacles. However, due to its finite capacity, the amount of data that a surveillance drone can store is limited. These collected data contains features or anomalies on a particular region and needs to be delivered properly.

Delivery drone $D$ moves toward to the surveillance drones to deliver the collected data. The two drones communicate using Wi-Fi Direct when it gets close. The transmission rate is 250Mbps, transmission distance is up to 200m, and it supports 1:N connection with devices at the same time. Note that data transmission can vary depending on the delivery drone's battery capacity, specification, current location, etc. In our auction, the delivery drones bid privately, strategically based on their calculated valuation.
In short, surveillance drone in remote and insecure area collects monitoring data in real-time and wants it to be processed. Delivery drones try to provide assistance in delivery as much as possible while in the air to earn revenue from third-party operators and deliver data to their area in charge. Thus, the delivery drones flying in the path naturally compete to deliver sensing data from the Surveillance drone.
It is shown in Fig.~\ref{fig:architecture} and their specification is in Tab.~\ref{tab:drone specification}. The operation of two types of the drone can be modeled as following two subsections.

\subsection{Drone Models}\label{sec:drone model}
\subsubsection{Surveillance Drone Model}
Surveillance drone has its own data buffer size $B$, and data flow can be formulated with a queue. The surveillance drone stores image from the mounted camera every time step, and a certain amount is processed by the delivery drone. In this paper, the battery of surveillance drone is not considered, assuming that the nearby ground charging tower covers the surveillance drone. Suppose that $S$ is the set of delivery drones that can participate in an auction of one-time steps, and let's denote each delivery drone as $d_i, \forall_{i} \in \{1,...,|S|\}$. In following~\eqref{eq:queue}, $Q(t)$ is the current queue size in storage, $\alpha_{i}(t)$ is the amount of single delivery drone can take, and $\lambda(t)$ is the size of stacking surveillance image in every time t. The amount of the data leaving the queue depends on the surveillance drone's request, i.e.,
\begin{eqnarray}
    Q(t+1) &=& Q(t) + \sum_{\forall d_{i}\in\mathcal{S}} (1-I_{i}^{S})\cdot \alpha_{i}(t) + \lambda(t), 
\label{eq:queue} \\
\text{where }        I &=&
    \begin{cases}
    0, & \text{scheduled}, \\
    1, & \text{otherwise}.
    \end{cases}
\end{eqnarray}

Surveillance drone hands over data to the selected delivey drone and acts in a way to maximize the profits as much as possible during the process.
For the rotary-wing drone, the hovering power consumption $P_{h}$ can be represented as the sum of the power $P_o$ needed to turn the blade around (rotor) and the power $P_i$ needed to lift the weight of the drone, i.e.,
\begin{equation}
    P_{h} = \underbrace{\frac{\delta}{8}\rho sA\Omega^{3}R^{3}}_{P_{o}} + \underbrace{(1+k)\frac{W^{3/2}}{\sqrt{2\rho A}}}_{P_{i}},
\end{equation}
where the parameters in this equation are summarized in Tab.~\ref{tab:parameters}~\cite{pugliese2016modelling,zorbas2016optimal}.

\subsubsection{Delivery Drone Model}
In the case of a delivery drone, the battery capacity determines its performance. The delivery drone has two modes of flight: 1) hovering along the path until it matches with the surveillance drone through auction, and 2) flying between two points for data delivery.
In other words, the energy expenditure of delivery drones is the sum of hovering and traveling power consumption.
The communication related energy is used for various communication functions such as signal transmission, computation, and signal processing. Typically, communication-related energy is not considered due to its relatively small value~\cite{fotouhi2019survey,zhan2017energy,zeng2019energy}. The energy consumption for time $T$ with speed $V$ can be formulated as follows where the value depend on several factors such as weight, air density, rotor disc area, blade angular velocity and etc as given in Tab.~\ref{tab:parameters}~\cite{pugliese2016modelling,zorbas2016optimal}, i.e.,

\begin{align}
    E = T \bigg[ &P_{0}\bigg(1 + \frac{3V^{2}}{U^{2}_{tip}}\bigg) + 
     P_{i}\bigg(\sqrt{1+\frac{V^{4}}{4v_{0}^{4}}} -\frac{V^{2}}{2v_{0}^{2}}\bigg)^{1/2} \nonumber\\
     &+ \frac{1}{2}d_{0}\rho sAV^{3} \bigg].
    \label{eq:energy}
\end{align}

The model for a delivery drone in this paper is DIJ Mavic 2, and its specification is shown in the following Tab.~\ref{tab:drone specification}. The amount of energy can be calculated with the specification parameters. Delivery drones make decisions whether to join the auction in consideration of the amount of energy with the energy model. 

\begin{table}[b]
\small
\centering
\caption{Drone Specification}
\begin{tabular}[t]{lcc}
\toprule[1.0pt]
& Surveillance drone & Delivery drone \\
\midrule[0.5pt]
Model & Phantom4 PRO & Mavic 2 \\
Size & 1\,ft\,(diagonal) & 322$\times$224$\times$84\,mm \\
Weight & 1388\,g & 907\,g \\
Speed (max) & 72\,km/h & 72\,km/h \\
Flight time (max) & 30\,min & 31\,min \\
Battery capacity & 5870\,mAh & 2970\,mAh \\
\bottomrule[1.0pt]
\end{tabular}
\label{tab:drone specification}
\end{table}


\subsubsection{Mobility Planning for Delivery Drones}

\begin{figure}[t]\centering
    \includegraphics[width=1\columnwidth]{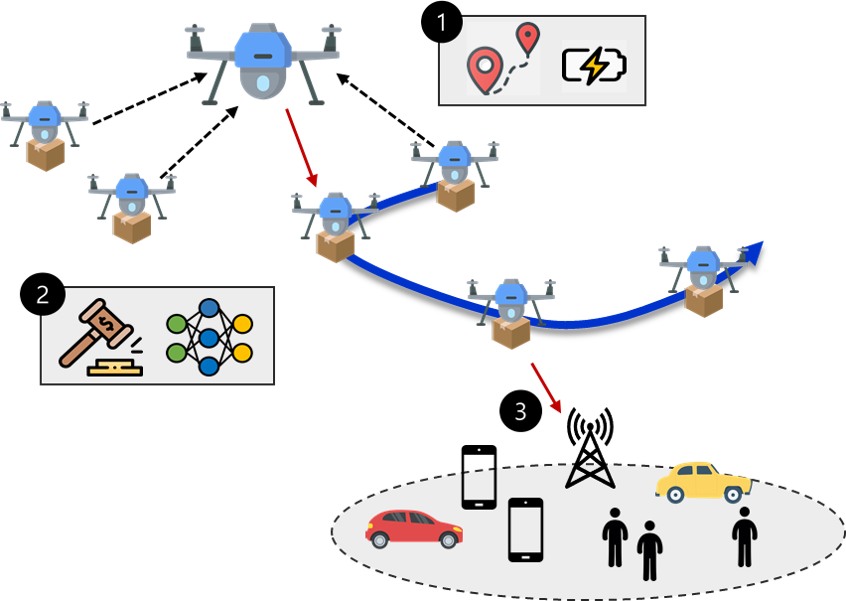}
    \caption{Delivery Drone Trajectory.}
    \label{fig:route}
\end{figure}

This section presents the drone behavior while carrying out the delivery mission and the movement can be divided into three steps as shown in Fig.~\ref{fig:route}. The first step is to determine whether delivery drone can perform a given task as described in Algorithm~\ref{alg:candidate}.
Surveillance drone in need of data transmission broadcasts the maximum allowed latency $T$ and amount of required data $D$ when requesting data to nearby delivery drones. Once the data amount is known, delivery drones can calculate the time $t_{\textit{transfer}}$ for data delivery based on the Wi-Fi direct where the transmission rate is $250$\,Mbps. Through the difference between the maximum latency $T$ and $t_{transfer}$, maximum flight time $ t_{flight}$ is computable. We assume that delivery drones know the total flying distance $l_{i}$ from their current location to the final destination. The minimum required flight velocity $v_{\min}$ is obtained via distance $l_{i}$ and flight time $t_{fly}$. Next, using~\eqref{eq:energy}, we can calculate the amount of energy consumed with constant flying speed $v_{\min}$. Each delivery drone can determine whether they participate in the auction or not by comparing the remaining energy and the amount of energy required. Among the eligible delivery drones, the one selected as winner fly to the area where it can communicate with the surveillance. The winner drone receives the collected data over WiFi-direct.
In the second step, the delivery drone flies all the way to the BS coverage. We assume that the drone is in a forward flight mode with constant speed on the journey. Refer to the subsection~\ref{sec:drone model} for energy consumption along the path.
In the final step, a delivery drone that reaches the BS transfers the collected data. Then the BS distributes data to mobile devices in coverage. Alternatively, it can also be passed it to a big data platform, which enables data analysis through distributed data collections. After completing the mission, the delivery drone charges battery and start re-positioning.
\begin{table}[t]\centering
\small
\caption{Notations}
\begin{tabular}{@{}lr@{}}
\toprule[1.0pt]
Notation & Value \\
\midrule[0.5pt]
Aircraft weight, $N$ & $8$ \\
Rotor radius, $R$ & $0.4$ \\
Rotor disc area, $A = \pi R^{2}$ & $0.503$ \\
Number of blades, $b$ & 4\\
Rotor solodity, $s = \frac{0.0157b}{\phi R}$  & 0.05\\
Blade angular velocity, $\Omega$ & 300\\
Tip speed of the rotor blade, $U_{tip}=\Omega R^2$ & 120\\
Fuselage drag ratio, $d_{0}=\frac{0.0151}{sA}$ & 0.6\\
Air density, $\rho$ & 1.225\\
Mean rotor-induced velocity in hovering, $v_{0}=\sqrt{\frac{W}{s\rho A}}$ & 2.54\\
Profile drag coefficient, $\delta$ & 0.012\\
Incremental correction factor to induced power, $k$ & 0.1\\
\bottomrule[1.0pt]
\end{tabular}
\label{tab:parameters}
\end{table}

\section{Learning-based Optimal Auction for Autonomous Aerial Delivery}\label{sec:5}

\subsection{Auction Design Concepts}

In a single-item mechanism $M = (g(\textbf{b}), p(\textbf{b}))$ with a set of \textit{N} of $n$ bidders consists of an allocation and payment rule. Allocation rule choose a feasible allocation $\textbf{g(b)} \in X \subseteq \textit{R}^n$ as a function of the bids, which is $\sum{g_i(b)} \leq  1$. Payment rule choose payments  $\textbf{p(b)} \in \textit{R}^n$ as a function of the bids. And Bidder $i$ has utility $u_{i}(\textbf{b}) = v_{i} \cdot g_{i}(\textbf{b}) - p_{i}(\textbf{b})$. In our auction settings, allocation and payment rule follows a standard SPA and only the concept of Myerson's virtual valuation is added. To truthfully allocate items, the mechanism must deter the presence of malicious bidders . Here are several desirable properties that a truthful mechanism should hold. 

\begin{defi}[Individual Rationality (\textsf{IR})]
  A \textit{truthful mechanism $M = (g(\textbf{b}), p(\textbf{b}))$ is individually rational for all bidders, if their utilities are more than 0}.
\end{defi}

\begin{equation}
     U_i(b) \geq 0, \forall i \in N 
\end{equation}

\begin{defi}[Incentive Compatibility (\textsf{IC})]
  A \textit{truthful mechanism $g(\textbf{b}), p(\textbf{b})$ is incentive compatible if no requester can improve its utility by misreporting its bid}.
\end{defi}

\begin{equation}
    U_i(b_i, b_{-i}) \geq U_i(\hat{b}_{i}, b_{-i}), \forall \hat{b_i} \in \eta(i), \forall i \in N
\end{equation}

\begin{defi}[Budget Balance (\textsf{BB})]
  A \textit{truthful mechanism $g(\textbf{b}), p(\textbf{b})$ is individually rational for all bidders, if their utilities are more than 0}.
\end{defi}

\begin{equation}
    p_i(b) \leq B_i, \forall i \in N
\end{equation}

With the allocation, payment rule and the monotonic transform function, the objective of maximizing the surveillance drone's revenue can be achieved. The computed revenue for winning bidder is $R(g(\textbf{b}), p(\textbf{b}))$, which can be formulated as,
\begin{equation}
    R(g(\textbf{b}), p(\textbf{b})) = E_{\textbf{b}\sim F} \left\{\sum_{\textit{i}\in\textit{N}} {(p_i(\textbf{b})-c)\cdot g_i(\textbf{b})} \right\}
\end{equation}

where $c$ is the processing cost of surveillance drone for transmitting a unit of data to winner drone. And we assume that every bidder's private valuations follows the same distribution as in~\eqref{eq:distribution}. Therefore, our auction problem can be formulated as,

\begin{eqnarray}
\max & & R(g(\textbf(b)), p(\textbf(b)))\\
\textrm{s.t.} & & \textsf{IR}: U_i(b) \geq 0, \forall i \in N \\
  & &\textsf{IC}: U_i(b_i, b_{-i}) \geq U_i(\hat{b}_{i}, b_{-i}), \forall \hat{b_i} \in \eta(i), \forall i \in N   \\
  & &\textsf{BB}: p_i(b) \leq B_i, \forall i \in N
\end{eqnarray}
where this optimization program satisfies when $\phi$ is a strictly monotone.

\subsection{Auction Design for Delivery Drone Scheduling}
\subsubsection{Auction-based Delivery Drone Scheduling Process}
\begin{figure}[t]\centering
    \includegraphics[width=1\columnwidth]{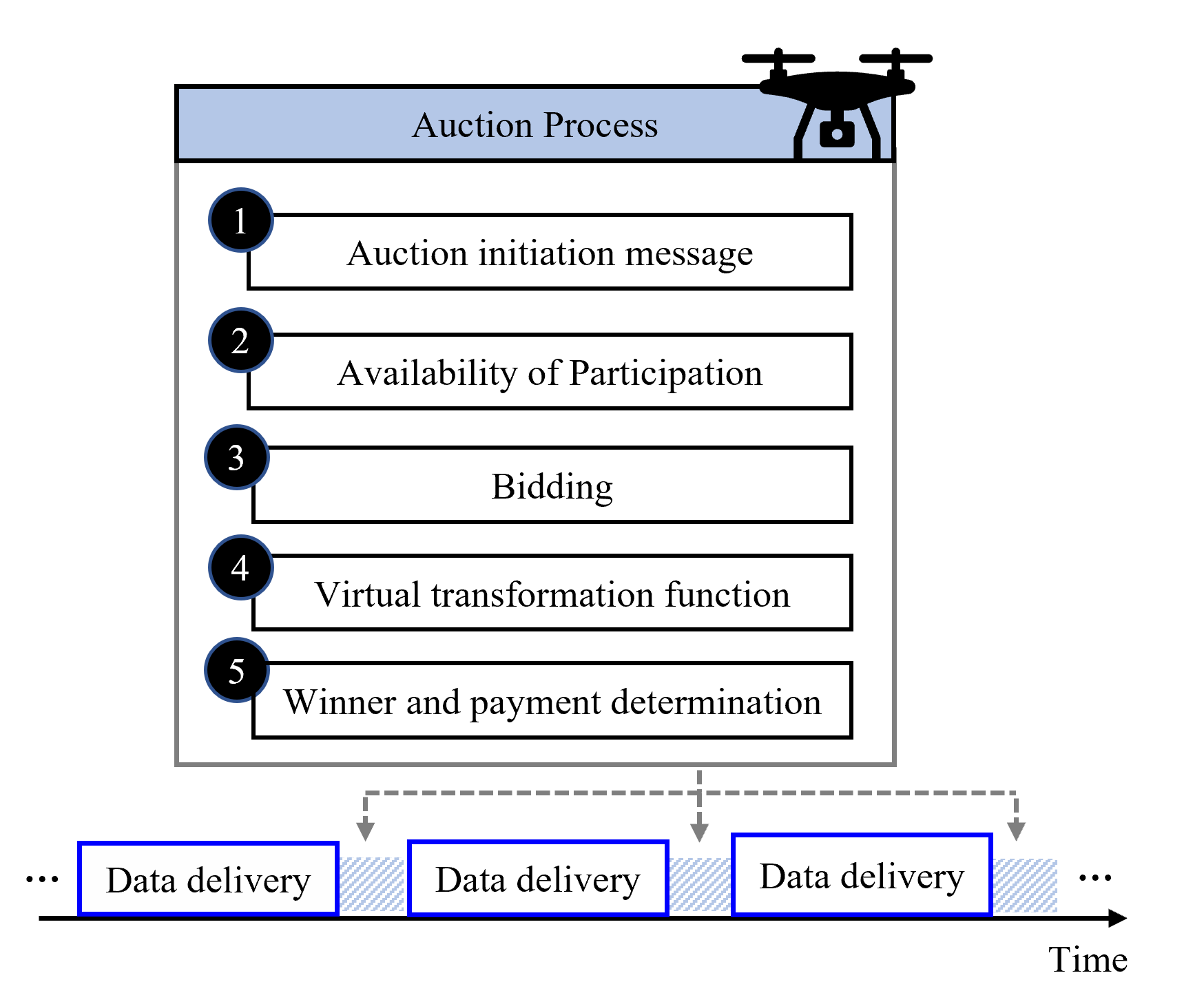}
    \caption{Auction Computation Process}
    \label{fig:auctionprocess}
\end{figure}

To start the auction process in Fig.~\ref{fig:auctionprocess}, a surveillance drone in need broadcasts an auction start message and its delivery conditions, i.e., minimum data amount $D$, maximum delay time $T$. The delivery drone in coverage (1) receives the auction request message and delivery conditions. (2) In the participation decision stage, each delivery drone that receives the start message first determines whether it is available to attend the auction.
A comparison between the minimum required energy for data transport and the residual energy for each individual drone is needed.
(3) The participating delivery drones bid based on their valuation to increase their utilities. 
(4) Then the surveillance drone collects all bids from the delivery drone and which are in transform form. (5) The drone with the highest allocation probability becomes a winner and pays final the determined payment and is detailed in Sec.~\ref{sec:auction}.

\begin{algorithm}
  \caption{Candidate Selection}
  \label{alg:candidate}
  \hspace*{\algorithmicindent} \textbf{Input:} $D$, $T$, $location$, $remain$, $i= 1, 2, \cdots, n$ \\
  \hspace*{\algorithmicindent} \textbf{Output:} $Candidate$ list, $bid$ list
  \begin{algorithmic}[1]
    \STATE $candidate \gets  [\ ] $
    \STATE $t_{transfer} \gets D \times 8 / 250$Mbps
    \STATE $t_{fly} \gets T - t_{transfer}$
    \FOR{\texttt{$i \gets 1 \ to \ N$}} 
    \STATE \texttt{distance} $\gets getDistance(d_{i}, s_{i}, b_{i})$    \COMMENT{location of drones $d_{i}$, $s_{i}$ and base station $b_{i}$}
    \STATE \texttt{$v_{min}$} $\gets$ \texttt{distance} $/t_{fly}$
    \STATE \texttt{energy} $\gets getEnergy$(\texttt{$v_{min}$}) \COMMENT{Eq.~\ref{eq:energy}}
    \IF{\texttt{energy} $\ >= \ $\texttt{remain}}
        \STATE Append to $candidate$
    \ENDIF
    \ENDFOR
    \FOR {each item $i$ in $candidate$}
    \STATE $b_{i} = \frac{d_{i}}{p_{i}}$ 
    \ENDFOR
    \RETURN $candidate$, $bid$
    \STATE System Initialization
  \end{algorithmic}
\end{algorithm}

\subsubsection{Individual Bid Valuation}
Drones independently determine the bid value according to their valuations. The valuation of drones can be modeled with ground demand in area as $d_i$ and the sacrificed energy ratio as $p_i$ in~\eqref{eq:valuation}.
\begin{equation}
    v_{i} = \frac{d_{i}}{p_{i}}.
    \label{eq:valuation}
\end{equation}

Here, the demand of mobile devices in the base station coverage, which is the drone's final destination, can be denoted as $d_i$, and its value is in range 0 to 1. The delivery drones have the mission of providing information to the ground users in the area covered by the drone. The initial amount of individual energy can be denoted as $b_i$, the amount of total energy for the delivery mission as $e_i$, and the residual energy as $r_i$, i.e., $e_i = b_i + r_i$. Then the ratio of consumption energy to initial energy can be denoted as $p_i = \frac{b_i}{e_i}$.
It can be calculated with the drone energy model \cite{zeng2019energy}, in consideration of minimum data amount, maximum delay time, distance, and drone specification. In general, when the value of $d_i$ is larger, the drones are willing to join the auction and pay cost for the chance to match with the surveillance drone. On the contrary, when the $p_i$ is large, the drones would be less incentive to join the auction.
Bidders' valuation profile is drawn from a distribution $f_V(v)$.
Thus, the distribution $f_V(v)$ can be determined based on the distribution of $d_{i}$ and $p_{i}$ denoted as $f_{D}(d)$ and $f_{P}(p)$ respectively.

Due to the deficiency of prior knowledge, we assume the two variables $d$, $p$ are independent and uniformly distributed in range $d_{i} \sim U[d_{\min}, d_{\max}]$ and $p_{i} \sim U[p_{\min}, p_{\max}]$. To apply the Jacobian transformation, $v$ is set as $v = \frac{d}{p}$ and $z$ is $z = p$, then $d$ and $p$ is $d = p \times v$, $p = z$ respectively. The Jacobian matrix is as follows and determinant $J$ is equal to $z$, i.e.,

\begin{align}
J &=
\begin{vmatrix}
\frac{\partial_{d}}{\partial_{v}} & \frac{\partial_{d}}{\partial_{z}} \\ 
\frac{\partial_{p}}{\partial_{v}} & \frac{\partial_{p}}{\partial_{z}}  \notag
\end{vmatrix}
=
\begin{vmatrix}
z & v \\ 
0 & 1
\end{vmatrix}
= z.
\\ \notag
\end{align}

Given that the $d$ and $p$ follow uniform distribution, the joint distribution $f_{V, Z}(v, z)$ can be obtained as,
\begin{eqnarray}
    f_{V, Z}(v, z) &=& f_{D,P}(d(v, z), p(v, u))|J(V,Z)| \\
    &=& f_{d}(v,z)f_{p}(z)|z|\\
    &=& \frac{1}{(d_{\max}-d_{\min})(p_{\max}-p_{\min})}|z|.
\end{eqnarray}

As a sequence, the distribution of $v$, i.e., $f_V(v)$, which is the marginal function can be derived as,
\begin{eqnarray}
    f_{V}(v) &=& \int f_{V,U}(v, u) \,du \\
    &=& \int_{p_{\min}}^{p_{\max}} \frac{1}{(d_{\max}-d_{\min})(p_{\max}-p_{\min})}|u| \,du \\
    &=& \frac{p_{\max}+p_{\min}}{2(d_{\max}-d_{\min})}.
    \label{eq:distribution}
\end{eqnarray}

Therefore, each drone submits bid according to its private value, where it is in between $v \sim [p_{\min}/d_{\max};p_{\max}/d_{\min}]$.

When delivery drones compete for data delivery, there is a possibility of malicious drone bids higher than its value. Our auction needs to let the participants act truthfully to ensure system stability and achieve revenue-optimal in the same time. Since Myerson presents provable analytical results for single item auction which can optimize the auctioneer revenue where each buyer has its own private valuation of the resource while guarantees truthfulness \cite{myerson1981optimal}, we used a variant of Myerson auction using deep learning.

\subsection{Neural Myerson Auction for Optimal Delivery}\label{sec:auction}

This section presents how deep learning-based auction maximizes the expected revenue of surveillance drone while guaranteeing truthfulness and revenue-optimal. The monotonic network is used for random sampling for approximating pseudo-optimal revenue values. In addition, allocation networks and payment networks are for determining the winner drone and the payment, respectively. Detailed neural architectures for deep learning to solve our proposed auction-based problems are organized in Algorithm~\ref{alg:deep} and presented as following subsections, i.e., monotonic networks (refer to Sec.~\ref{sec:4-1}), allocation networks (refer to Sec.~\ref{sec:4-2}), and payment networks (refer to Sec.~\ref{sec:4-3}) 

\begin{figure}\centering
    \includegraphics[width=1\columnwidth]{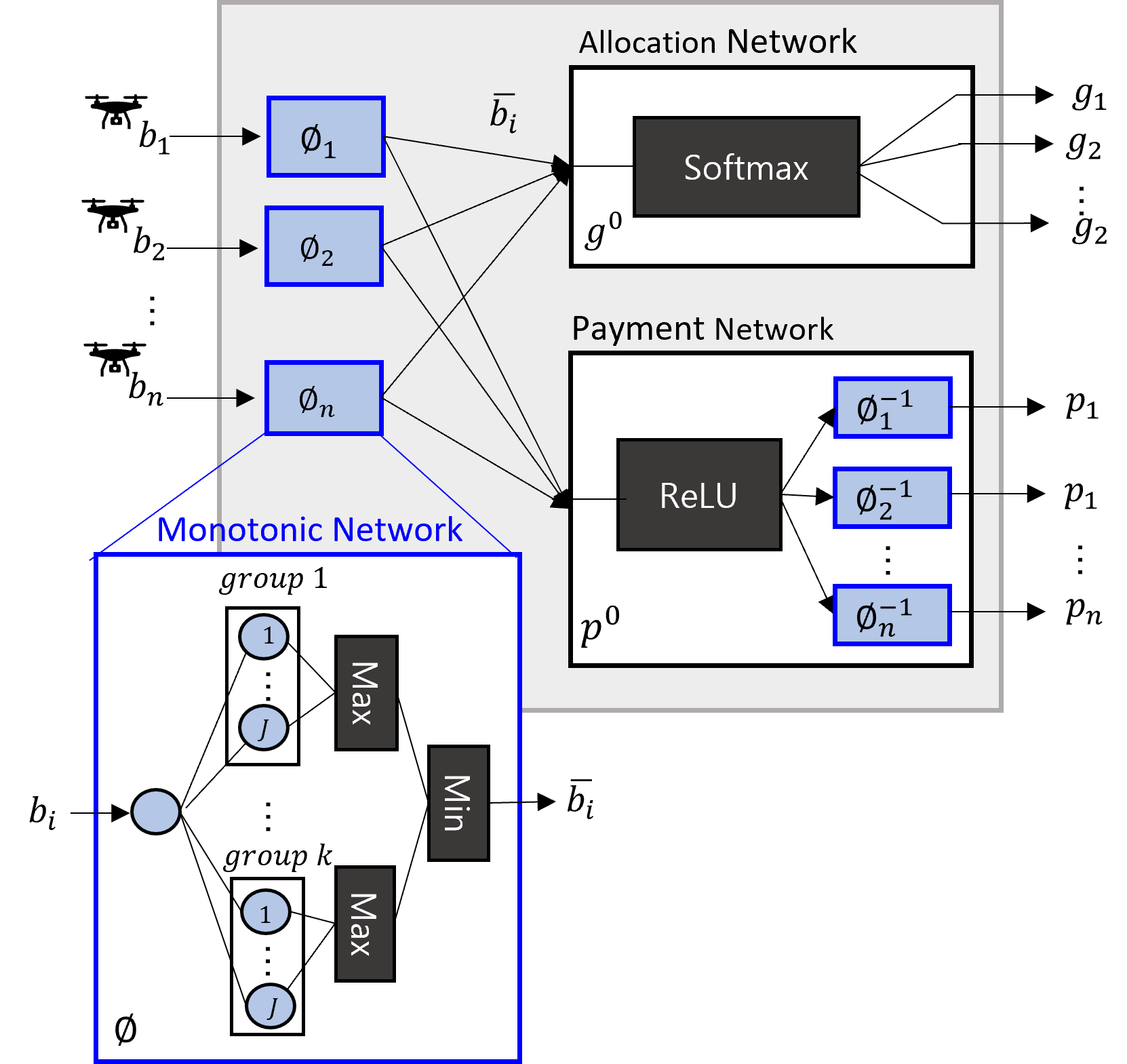}
    \caption{Deep learning auction framework}
    \label{fig:framework}
\end{figure}

\subsubsection{Virtual Valuation Function}\label{sec:4-1}
Our virtual valuation function in auction network is denoted as $\phi_i$ and takes the role of virtual valuation in Myerson auction~\cite{luong2018optimal}. The input bids $b_i$ of delivery drones are transformed to $\bar{b_i}$ as it passes the monotonic network which consists max/min operations over several linear functions. 
Monotonic network $\phi_i$ uses $K$ groups of $J$ linear functions and is defined as follows~\cite{luong2018optimal}.

\subsubsection{Winner Determination Function}\label{sec:4-2}

The SPA allocation network maps the surveillance drone and delivery drone with the highest non-zero transform bid. The layer of softmax draws the allocation probabilities as an output with the transformed bids $\bar{b_i}$ and dummy input $\bar{b_{N+1}} = 0$. Semantically, the softmax is used for taking the maximum value. The allocation network with softmax can be represented as follows,

\begin{eqnarray}
g_i(\bar{b}) &=& \textsf{softmax}_i(\bar{b}_{1},...,\bar{b}_{N+1} ; k) \\
&=& \frac{e^{k\bar{b}_i}}{\sum_{j=1}^{N+1}e^{k\bar{b}_j}}, \forall i \in N
\label{eq:3}
\end{eqnarray}
where $k$ is a parameter of softmax function and it determines the quality of the approximation~\cite{tvt201905shin,luong2018optimal}.
\subsubsection{Payment Function}\label{sec:4-3}
The payment network determines the final payment to the winner delivery drone. Payment network uses a ReLU activation function as follows to make the payment non-negative,
\begin{equation}
p_i^{0}(\bar{b}) = ReLU(\max_{\forall j \neq i}\bar{b_j}), \forall i \in N.
\label{eq:4}
\end{equation}

Finally, the final payment of the winner delivery drone to surveillance drone can be calculated as follows,
\begin{equation}
p_i = \phi_i^{-1}\left(p^{0}_i(\bar{b})\right).
\end{equation}
\subsubsection{Neural Network Training}\label{sec:4-4}
Neural architecture trains parameters $w^{i}_{kj}$ and $\beta^{i}_{kj}$ with the valuation profiles as the training set and minimize the loss function. Here, we defined the loss function as the negative revenue in Myerson auction.
The loss function $\hat{R}$ is defined as follows,
\begin{equation}
\hat{R}(w,\beta) =-\sum_{i=1}^{N}\nolimits  g_i^{(w,\beta)}(v^{s})p_i^{(w, \beta)}(v^{s}).
\label{eq:5}
\end{equation}

The results of allocation networks and payment networks are used for training parameters, and we used the stochastic gradient descent optimizer to train the loss function $\hat{R}$.

\begin{algorithm}[t]
   \caption{Deep Learning-Based Auction Algorithm}
   \label{alg:deep}
\begin{algorithmic}[1]
   \STATE {\bfseries Input:} Candidate bid sets $\textbf{b}=(b_{1}, b_{2},...,b_{N})$ \\
   \STATE {\bfseries Output:} Allocation probability set $g_i=(g_{1}, g_{2},...,g_{N})$, payment set $p_i=(p_{1}, p_{2},...,p_{N})$
   \REPEAT 
   \STATE Compute \ $\phi_i(b_i) = \min_{\forall k \in K}\max_{\forall j \in J} \left(w^{i}_{kj}b_i + \beta^{i}_{kj}\right)$ ; 
   \STATE Compute \ $g_i(\bar{b}) = \frac{e^{k\bar{b}_i}}{\sum_{j=1}^{N+1}e^{k\bar{b}_j}}$ ;
   \STATE Compute $p_i^{0}(\bar{b}) = ReLU(\max_{\forall j \neq i}\bar{b_j})$ ;
   \STATE Compute\ $\phi^{-1}_i(y) = \max_{\forall k \in K}\min_{\forall j \in J} \left(w^{i}_{kj}\right)^{-1}\left(y-\beta^{i}_{kj}\right)$;
   \STATE Compute \ $\hat{R}(w,\beta) =-\sum_{i=1}^{N}g_i^{(w,\beta)}(v^{s})p_i^{(w, \beta)}(v^{s})$ ;
   \UNTIL{The loss function $\hat{R}(w,\beta)$ minimizes}
\end{algorithmic}
\label{al:auction algorithm}
\end{algorithm}

\subsubsection{Auction Properties}
In previous section, we define the truthful characteristics and the auction network consisting allocation rule $g$ and payment rule $p$. According to Myerson theorem, we can guarantee the truthful condition IR and IC.

\begin{theorem}[Myerson~\cite{myerson1981optimal}]
For single parameter environments, any set of strictly monotone functions $\phi_1, \phi_2,...,\phi_N$, an auction that assigns an item to the bidder with highest virtual valuation $\phi_i(v_i)$ and the payment is determined by the second highest virtual valuation is IR and IC.
\end{theorem}

Let the neural architecture that have K groups and the outputs are denoted by $t_1, t_2,...,t_R$. Let $h_r$ denote the number of hyperplanes within group $r, r=1,2,...,R$. The parameters of the hyperplanes are denoted by $\mathbf{w}_{(r,1)}, \mathbf{w}_{(r,2)},...,\mathbf{w}_{(r,h_r)}$, where the matrix of all weights and biases is denoted by $\mathbf{W}$. 
Then, the output at group $r$ is $t_r(x)=\displaystyle\min_{j}(\mathbf{w}_{(r,j)}\cdot \mathbf{x} + \theta_{(r,j)}), 1 \le j \le h_r$ and the final output is $\mathbf{O_x} = \displaystyle\max_{r}{\mathit{t_r(x)}}$~\cite{daniels2010monotone}, which is same as our virtual valuation network in~\ref{sec:4-1}. This network obeys increasing monotonicity when all weights in the first layer are constrained to be positive~\cite{daniels2010monotone}, which is the satisfied condition in our system.

\section{Performance Evaluation}\label{sec:6}
\subsection{Evaluation Setup}
\subsubsection{Simulation Environment}
\begin{figure}\centering
    \includegraphics[width=1\columnwidth]{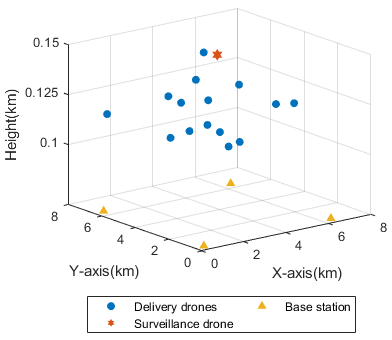}
    \caption{Example of drone deployment}
    \label{fig:bidder drones in 3d}
\end{figure}

For the simulation study, we placed four base stations are placed at the edge of $7$\,km $\times$ $7$\,km size map and surveillance drone was placed in the center and delivery drones were randomly placed at $150$ to $150$\,m high. Fig.~\ref{fig:bidder drones in 3d} shows an example of drone deployment in 3D space and their position is listed in Tab.~\ref{tab:drone location} checked with candidate availability. In this example, $15$ delivery drones exist around the surveillance drone and only 5 of them are available to attend the auction. Each delivery drones consider its energy and, total round-trip distance from the initial location to the base station via surveillance drone.

\begin{table}\centering
\caption{Locations}
\begin{tabular}{lcc}
\toprule[1.0pt]
 & Location & Availability \\
\midrule[0.75pt]
Surveillance drone & $(3.5000, 3.5000, 0.1500)$\\
\midrule[0.5pt]
Delivery drone $1$ & $(5.3891, 6.4843, 0.1018)$ & \checkmark\\
Delivery drone $2$ & $(1.0263, 6.9607, 0.1159)$ & \checkmark\\
Delivery drone $3$ & $(2.1797, 3.9923, 0.1275)$ & \checkmark\\
Delivery drone $4$ & $(5.4356, 4.6516, 0.1272)$ & \checkmark\\
Delivery drone $5$ & $(5.5366, 2.5679, 0.1233)$ & \checkmark\\
Delivery drone $6$ & $(3.9473, 3.9079, 0.1090)$ & \xmark\\
Delivery drone $7$ & $(1.0103, 3.1432, 0.1151)$ & \xmark\\
Delivery drone $8$ & $(5.8648, 1.8963, 0.1249)$ & \xmark\\
Delivery drone $9$ & $(5.1225, 4.2144, 0.1005)$ & \xmark\\
Delivery drone $10$ & $(2.2145, 3.5348, 0.1146)$ & \xmark\\
Delivery drone $11$ & $(1.6754, 1.7261, 0.1365)$ & \xmark\\
Delivery drone $12$ & $(4.2649, 3.7917, 0.1014)$ & \xmark\\
Delivery drone $13$ & $(5.5730, 6.9398, 0.1363)$ & \xmark\\
Delivery drone $14$ & $(3.8859, 5.2690, 0.1314)$ & \xmark\\
Delivery drone $15$ & $(1.8138, 4.2915, 0.1307)$ & \xmark\\
\midrule[0.5pt]
Base Station $1$ & $(6.5000, 0.5000, 0.0700)$\\
Base Station $2$ & $(0.5000, 0.5000, 0.0700)$\\
Base Station $3$ & $(0.5000, 6.5000, 0.0700)$\\
Base Station $4$ & $(6.5000, 6.5000, 0.0700)$\\
\bottomrule[1.0pt]
\end{tabular}
\label{tab:drone location}
\end{table}

\subsubsection{Algorithmic Setting - Bid Valuation}
We constructed the bid sets by randomly allocate the parameter value of the drone energy model. The initial energies of the drone battery are randomized in [2300, 2970]\,mAh with an output voltage of $7.6$\,V. For the stable drone operation, we calculated the actual available amount based on $80$ percent of the initial energies of the battery. As a sequence, we can derive the value $p_i$ which is the sacrificed energy ratio. And for the $d_i$ other component of private valuation, we randomly selected within [0, 1]. Tab.~\ref{tab:Bid value} shows the $10$ computed bidding samples of five actual participating delivery drones'. The values are various in the range between $0$ to $1$.

\begin{table}\centering
\caption{Data sample of delivery drones' bidding value}
\begin{tabular}{lccccc}
\toprule[1.0pt]
No. & Drone\, 1  & Drone\, 2 & Drone\, 3 & Drone\, 4 & drone\, 5 \\
\midrule[0.75pt]
$1$ & $0.6802$ & $0.4398$ & $0.8589$ & $0.7860$ & $0.9420$ \\
$2$ & $0.4552$ & $0.5123$ & $0.7315$ & $0.7600$ & $0.8045$ \\
$3$ & $0.5243$ & $0.5373$ & $0.7308$ & $0.8233$ & $0.8677$\\
$4$ & $0.6319$ & $0.7585$ & $0.8090$ & $0.8902$ & $0.9144$\\
$5$ & $0.4284$ & $0.4567$ & $0.5891$ & $0.7790$ & $0.8126$\\
\midrule[0.5pt]
$6$ & $0.3749$ & $0.6617$ & $0.7290$ & $0.8664$ & $0.9306$ \\
$7$ & $0.3347$ & $0.6277$ & $0.4597$ & $0.6433$ & $0.9502$ \\
$8$ & $0.3958$ & $0.6565$ & $0.7721$ & $0.8753$ & $0.9711$ \\
$9$ & $0.5070$ & $0.5135$ & $0.5687$ & $0.6221$ & $0.8643$ \\
$10$ & $0.1269$ & $0.4253$ & $0.5004$ & $0.8880$ & $0.9848$ \\
\bottomrule[1.0pt]
\end{tabular}
\label{tab:Bid value}
\end{table}

However, neither the prior knowledge of the valuation's distribution nor the real-world data is available in our situation. So the bid set is constructed based on the valuation of drones with the distribution assumptions. The auction simulation works through a randomly constructed bid set and infer results from it.

\subsection{Evaluation Results}
This section presents the DLA (deep learning-based optimal auction) for data delivery and the proposed deep learning based optimal auction compared with SPA as a baseline. The neural network runs on the pyTorch library. Evaluation was performed under where the numbers of delivery drones are $5$ and $7$ with the distribution of valuation $f_{V}(v)$ $\sim U[0.5,1]$ and the neural network has 5 groups and 3 linear functions. Overall 500 iterations were done with approximation quality $k$ is 1. The simulation parameters are organized in Tab.~\ref{tab:simulation parameters}.

\begin{table}\centering
\caption{Simulation Parameters}
\begin{tabular}{lcc}
\toprule[1.0pt]
Parameter & Value \\
\midrule[0.5pt]
Number of bidders (N) & 5, 7 \\
Number of groups (K) & 5 \\
Number of linear functions (J) & 3\\
Number of iterations & 500\\
Approximate quality k & 1\\
Distribution of valuation $f_{V}(v)$  & $\sim U[0.5,1]$\\
\bottomrule[1.0pt]
\end{tabular}
\label{tab:simulation parameters}
\end{table}

\begin{figure*}[ht!]
\centering
\setlength{\tabcolsep}{2pt}
 \renewcommand{\arraystretch}{0.2}
\begin{tabular}{cc}
\includegraphics[width=0.44\textwidth]{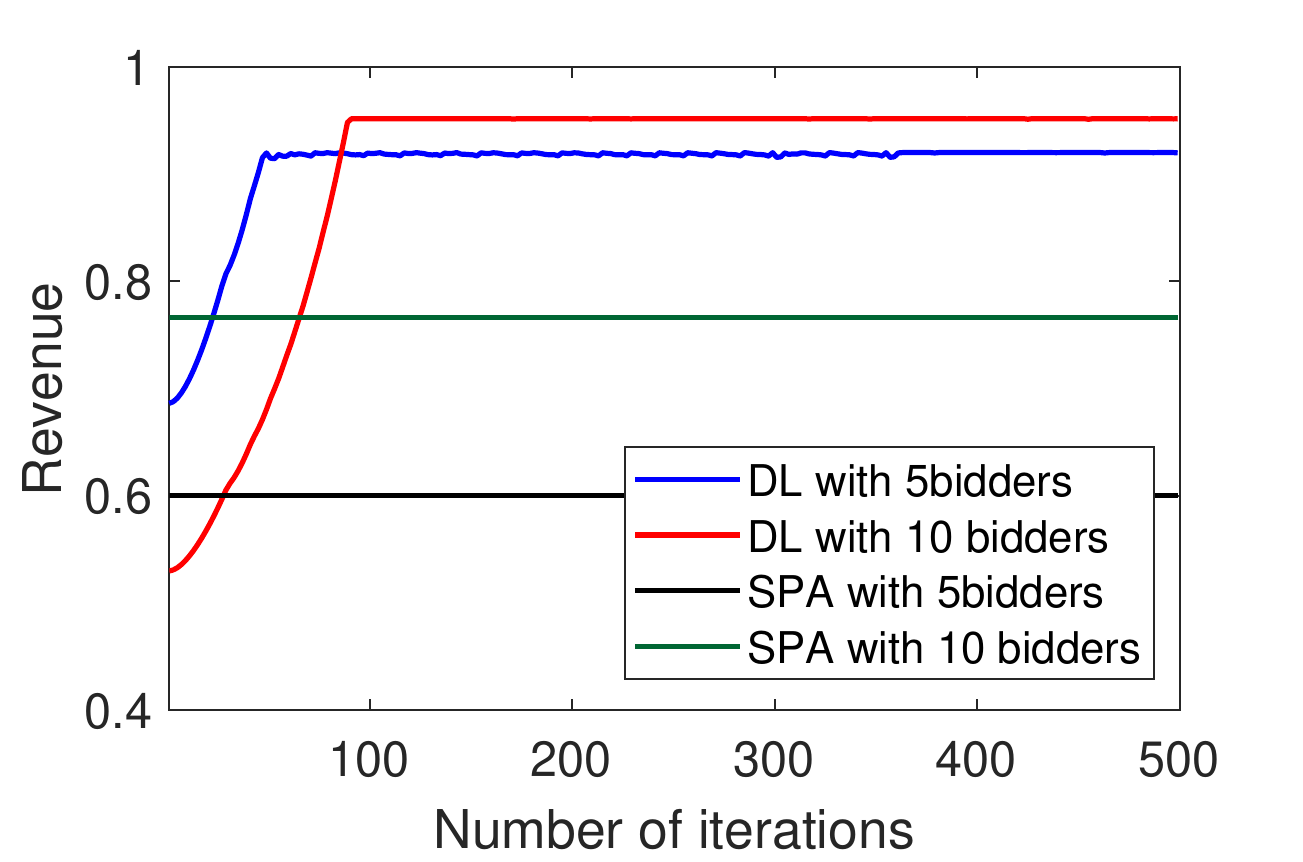} & \includegraphics[width=0.44\textwidth]{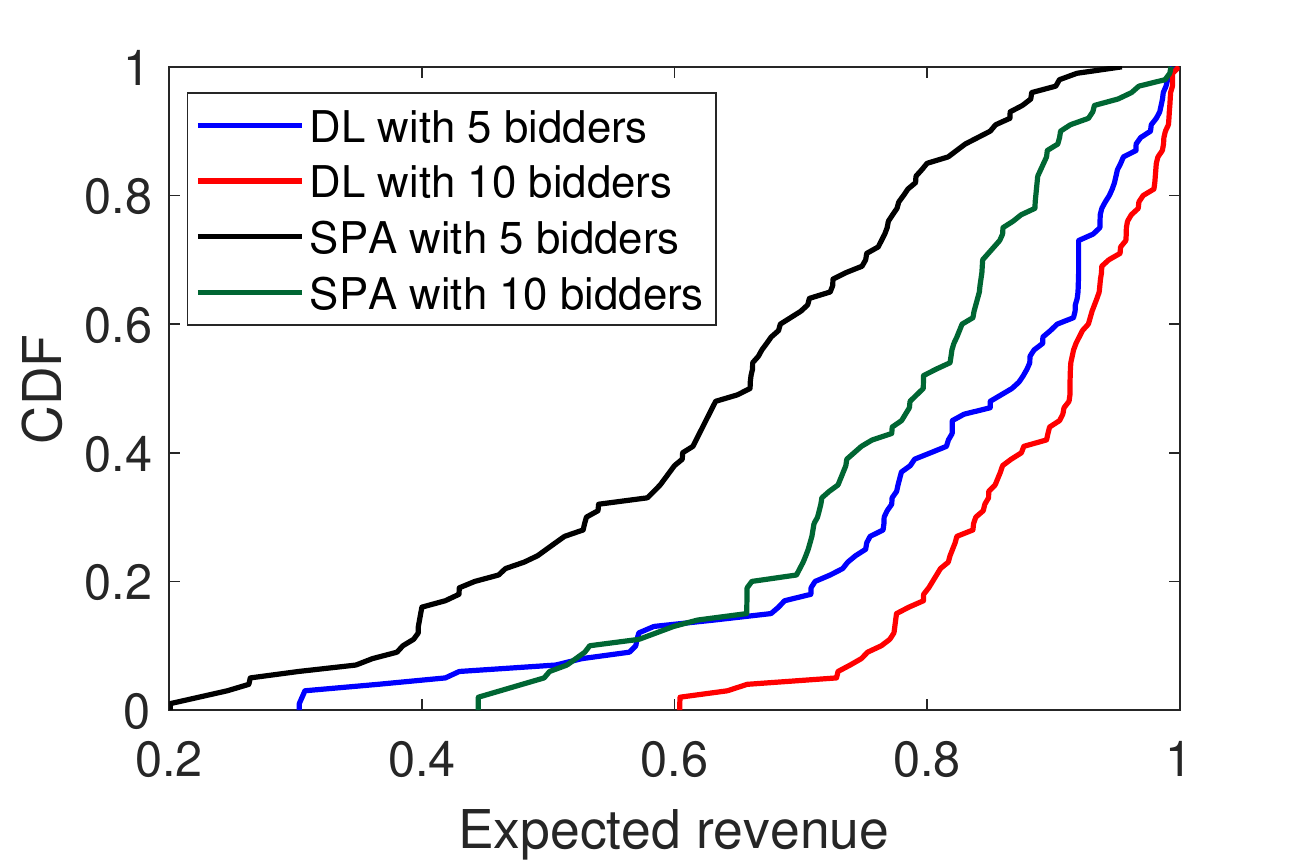}  \\
\tabularnewline
(a) Revenue of deep learning auction over iterations & (b) Cumulative distribution function (CDF) of revenue \\with 5 and 10 bidders. & derived from 100 experiment cases. \\
\tabularnewline
\end{tabular}
\vspace{0.1in}
\caption{Revenue comparison graph by auction method and number of participating bidders.}
\label{fig:revenue diff}
\end{figure*}

The results in Fig.~\ref{fig:revenue diff} shows revenue comparison between SPA and DLA in 4 different plottings. In Fig.~\ref{fig:revenue diff}(a), revenues comparison for 5 bidders and 10 bidders are shown over the iterations. The revenue obtained from the deep learning auction is higher than the baseline SPA for all cases. Fig.~\ref{fig:revenue diff}(b) represents the case where the participation of the five and ten bidder's in each of the DL and SPA process is shown as a CDF of their revenue. The graph shows that the revenue of the 10 bidders is higher than that of 5 bidders. It can be confirmed that our simulation reflects the obvious phenomenon that the bidding value increases as the number of competitors increases. The experiment results compare the top 25 percentile, 50 percentile, and 75 percentile value of CDF as listed in Tab.~\ref{tab:Figure 6. (b) statistic}. The mean value (i.e., 50\,\% of CDF) is $0.7055$, $0.7660$, $0.7357$, and $0.8480$, confirming the revenue increase through numerical values. 

\begin{figure*}[ht!]
\centering
\setlength{\tabcolsep}{2pt}
 \renewcommand{\arraystretch}{0.2}
\begin{tabular}{ccc}
\includegraphics[width=0.44\textwidth]{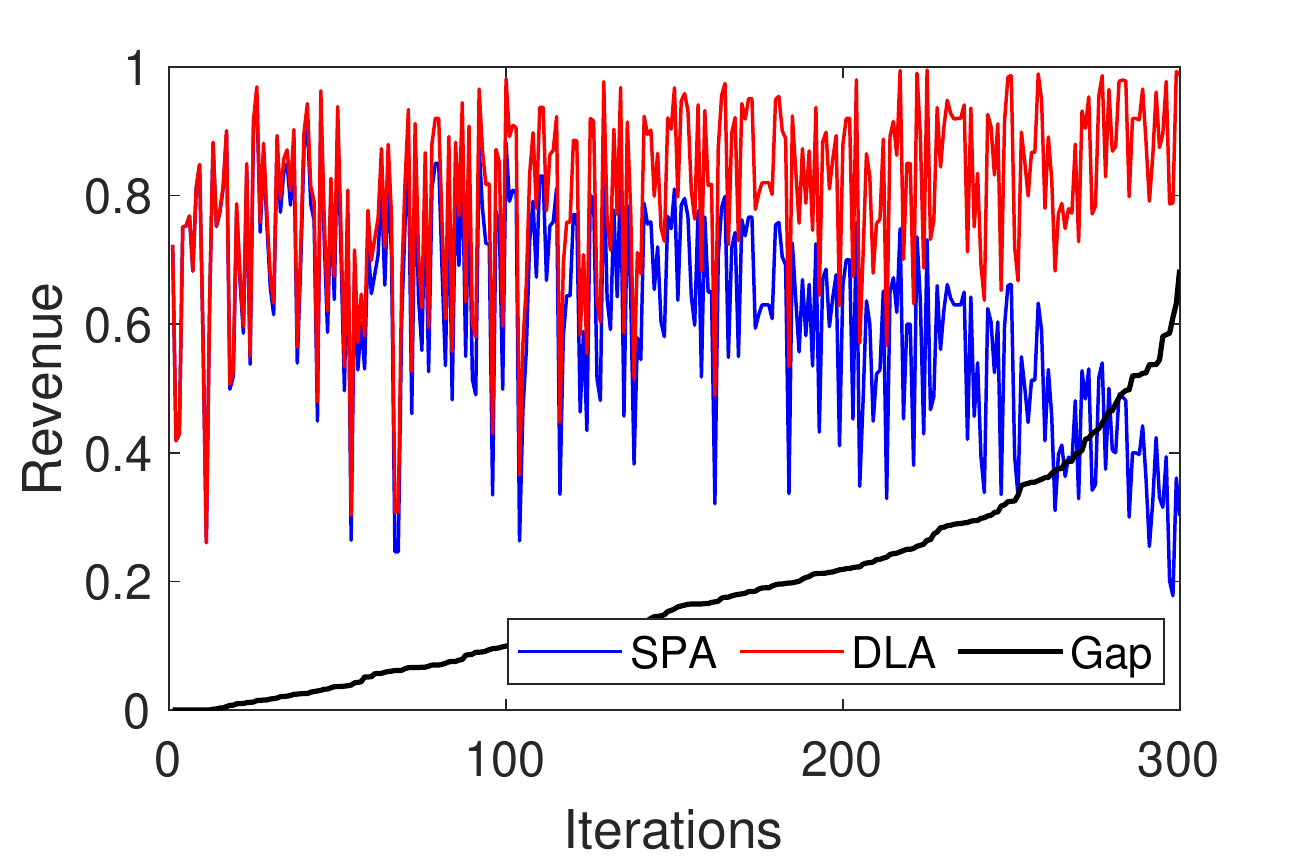} & 
\includegraphics[width=0.44\textwidth]{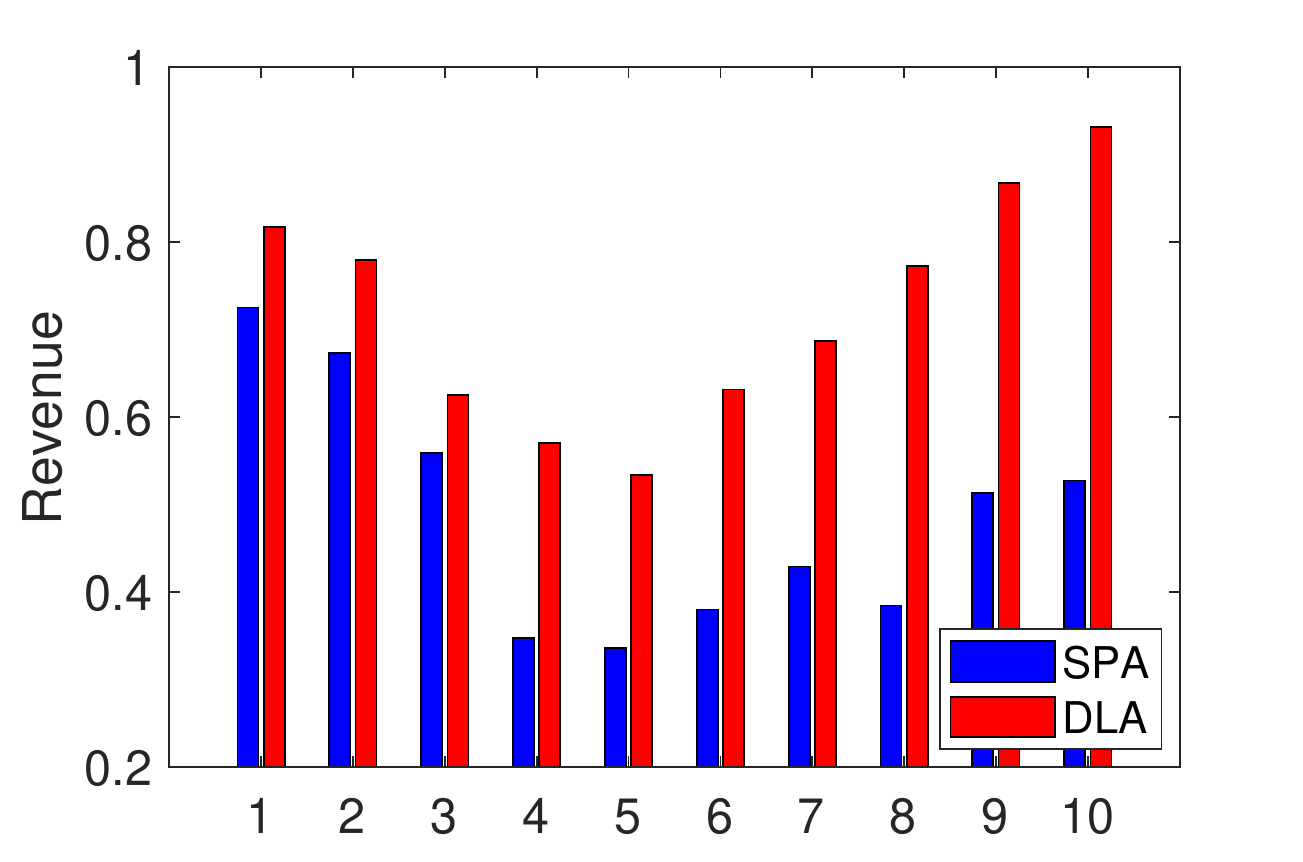}
\tabularnewline
(a) Revenue gap from 300 experiment cases  & (b) Revenue by auction method from \\
sorted in ascending order. & 10 experiment cases in barplot. 
\tabularnewline
\end{tabular}
\vspace{0.1in}
\caption{The gap between DLA and SPA}
\label{fig:revenue}
\end{figure*}

Fig.~\ref{fig:revenue}(a) show the 300 individual deep learning auction results. The revenue gap between spa and DLA is obtained for each iteration, and sorted in ascending order. That is, the corresponding graph shows the range of gaps that can occur over iterations. Overall, the revenue is improved, and the largest increase was up to 0.6. Fig.~\ref{fig:revenue}(b) shows the 10 cases of experiments in random order. The number on the X-axis represents the indices of individual cases. The result compares the revenue of surveillance drone via deep learning auction with SPA and FPA in barplot. Fig.~\ref{fig:revenue} confirms that the value is generally larger than the SPA and close enough to the FPA. 

\begin{table*}\centering
\caption{Statistic of CDF in figure 6. (b) }
\begin{tabular}{lcccc}
\toprule[1.0pt]
& SPA with 5 bidders & DLA with 5 bidders & SPA with 10 bidders & DLA with 10 bidders \\
\midrule[0.5pt]
25\,\% & $0.4987$ & $0.7514$ & $0.7059$ & $0.8222$ \\
50\,\% & $0.6598$ & $0.8672$ & $0.7970$ & $0.9132$ \\
75\,\% & $0.7684$ & $0.9368$ & $0.8601$ & $0.9578$ \\
\bottomrule[1.0pt]
\end{tabular}
\label{tab:Figure 6. (b) statistic}
\end{table*}

\begin{figure*}[ht!]
\centering
\setlength{\tabcolsep}{2pt}
 \renewcommand{\arraystretch}{0.2}
\begin{tabular}{cc}
\includegraphics[width=0.44\textwidth]{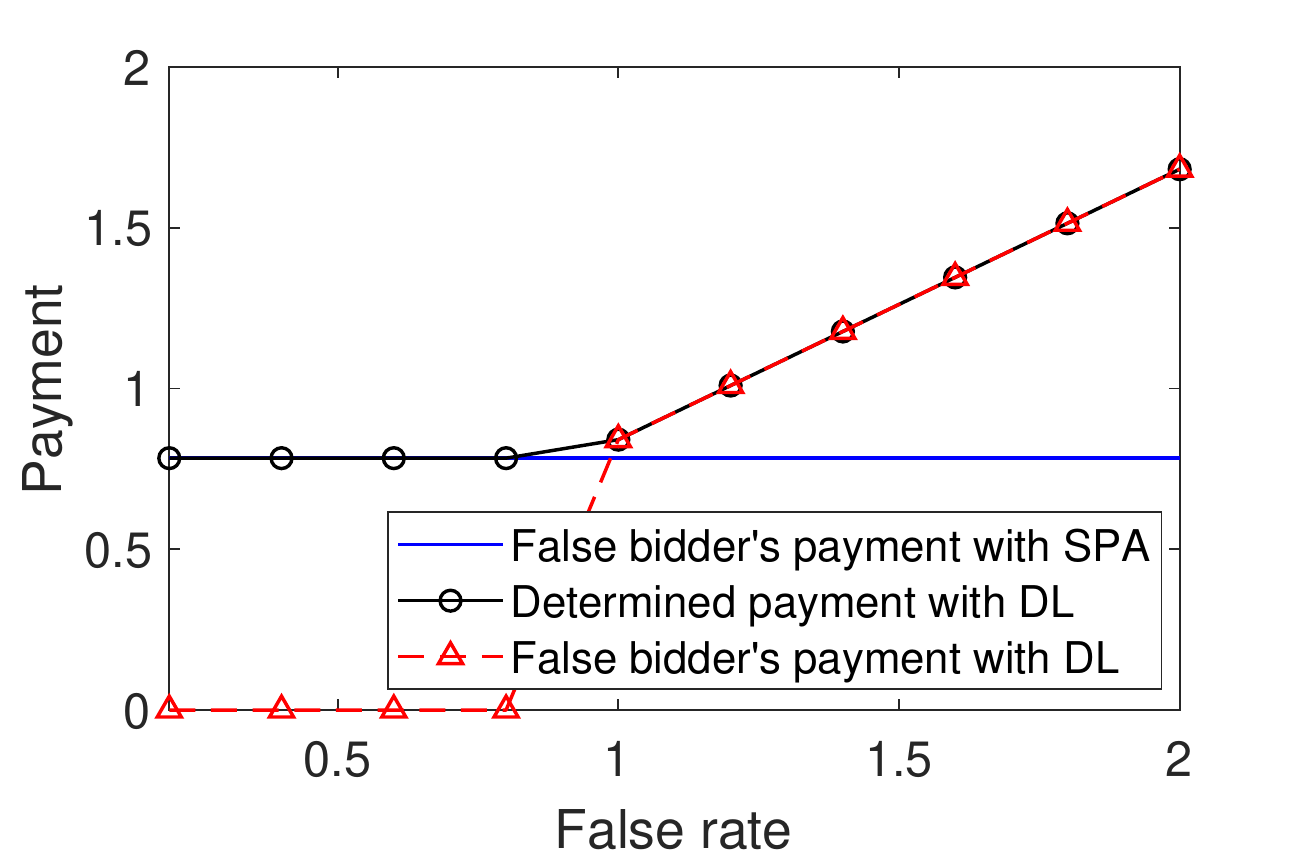} &
\includegraphics[width=0.44\textwidth]{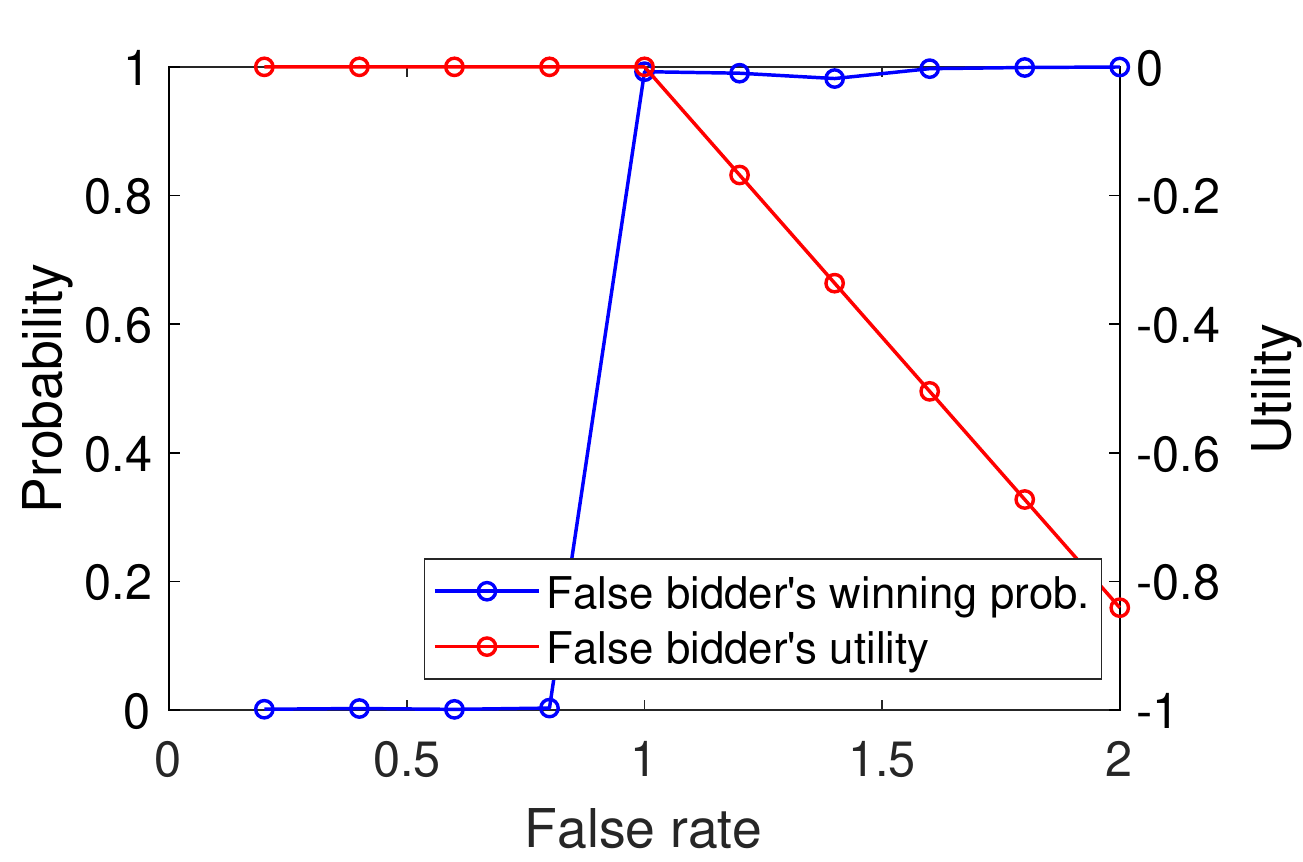}  \\
\tabularnewline
(a) Payment changes due to the false bidding. & (b) Correlation between probability and utility.
\tabularnewline
\end{tabular}
\vspace{0.1in}
\caption{The changes from the advent of false bidder}
\label{fig:falsebidding}
\end{figure*}
Fig.~\ref{fig:falsebidding} presents the changes from the advent of a false bidder drone. Fig.~\ref{fig:falsebidding}(a) shows the revenue changes due to the particular bidder's untruthful behavior. The experiment assumes the situation when five bidders bid for the item and one drone happens to false bid. The second-highest bid is 0.7832, and the truthful value of the malicious drone is 0.8408. We set the fake value by adjusting the false rate from 0.2 to 2.0. False rate is a multiplying value that indicates how much to adjust from the initial truthful value. The black line is the actual value taken by the winner drone in a deep learning auction, and the blue line is the fixed payment of the winner in the SPA. The payment change of the fake bidder over the false rate is in a red line. When the fake bidder drone submits a bid 0.2-0.8 times larger than the actual value, it has no chance to win the auction. When the fake bidder drone submits a bid 1.2-2.0 times larger than the actual value, it becomes the winner but overpays its valuation. In terms of the utility of individual drones, the loss leads the utility negative.
Therefore, there is no reason for a delivery drone to fake bid suffering the needless loss and shows our system prevents untruthful behavior.
Fig.~\ref{fig:falsebidding}(b) shows that bidders have no incentive to false bid in the same manner. The probability increases as the false rate increases, but the utility gradually decreases. With the false rate is equal to 1, the left and right sides of the graph show polar opposite characteristics. When the false rate is in 0.2-0.8, there is a little chance of winning, and when it is in 1.2-2.0, there is a high probability. However, when the false rate is 0.2-0.8, the false bidder cannot win the auction, thus the utility is zero. In addition, when it is 1.2-2.0, it has to pay more than its valuation, so it goes negative.

\section{Concluding Remarks}\label{sec:conclusion}
In this paper, an asynchronous data drone delivery is possible in aerial surveillance big data platforms. With the deep-learning auction, our platform achieves the initial objective of maximizing the revenue of the surveillance drone. The evaluation results confirm that the auction-based matching problem between the delivery drone and surveillance drone gives distinct revenue benefits compared to the traditional SPA. The results also give the reasonable inference that the participating drones are avoided from fake bidding. 

For future research directions, a deep learning-based multi-item auction can be considered to extend our proposed algorithm. The multi-item auction which processes data from multiple surveillance drones in distributed regions can be operated in realistic environment.


\vspace{10pt}
\bibliographystyle{IEEEtran}

\begin{IEEEbiography}[{\includegraphics[width=1in,height=1.25in,clip]{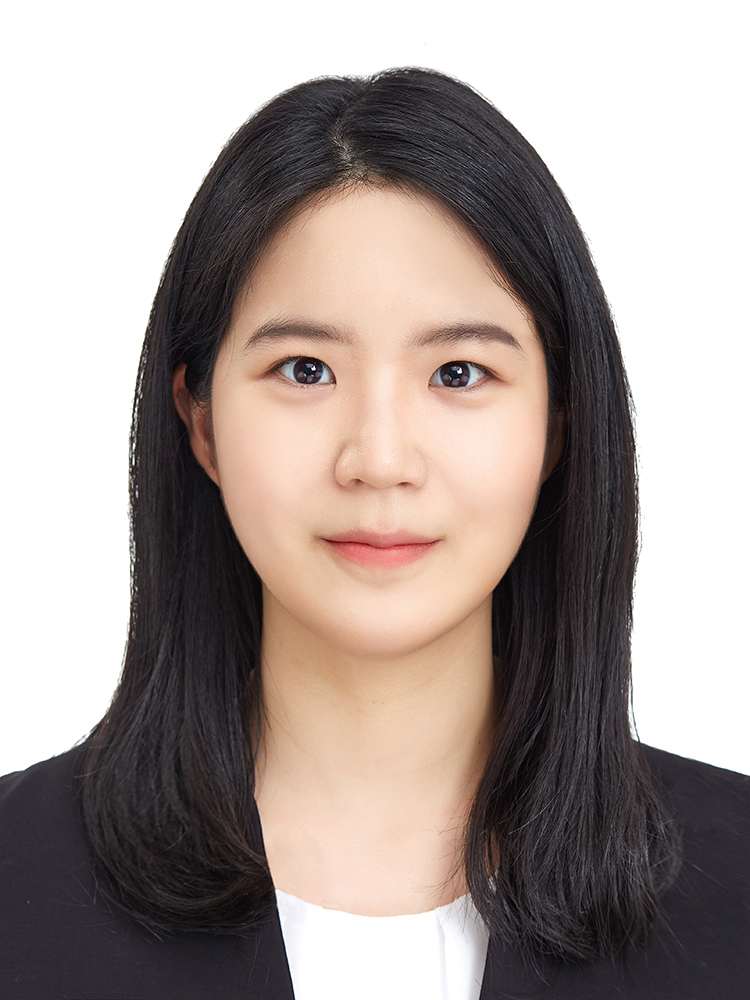}}]{Haemin Lee}
has been with Korea University, Seoul, Republic of Korea, since September 2020, and she is currently a Ph.D. candidate at the Department of Electrical and Computer Engineering, Seoul, Republic of Korea. She received the B.S. degree in statistics from Sookmyung Women's University, Seoul, Republic of Korea, in 2019.  Her research focuses include deep learning algorithms and their applications to communication networks.
\end{IEEEbiography}

\begin{IEEEbiography}[{\includegraphics[width=1in,height=1.25in,clip]{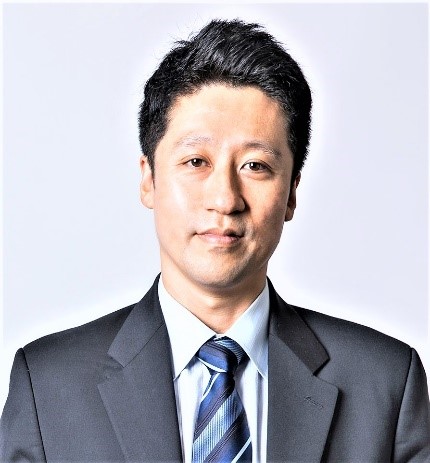}}]{Sean Kwon}
(S'09-M'14) received his Ph.D. degree from Georgia Institute of Technology in Atlanta, US on December 2013. Before that, Dr. Kwon received the M.Sc degree from the University of Southern California, Los Angeles, US in 2007; the B.Sc degree from Yonsei University, Seoul, South Korea in 2001.
He performed research at Intel's Next Generation and Standards Division in Communication and Devices Group during 2015 to 2017, where he contributed to 5G MIMO standards, the associated system design and patents. He has been Assistant Professor at California State University Long Beach; and Chief Director/Founder of Wireless Systems Evolution Laboratory (WiSE Lab) since 2017.

He also conducted postdoctoral research at Wireless Devices and Systems Group, University of Southern California in 2014 - 2015. He worked on CDMA common air interface focusing on layer-3 protocols at the R$\&$D Institute of Pantech co., Ltd, Seoul, South Korea in 2001 to 2004. He was involved in several projects such as a DARPA project; an US Army Research Lab project; and 6 mobile-station projects for Motorola and Sprint, which were successfully on the market. His current research interests are in 5G and beyond-5G wireless system/network design; satellite communications; polarization diversity and multiplexing; body area network such as wearable computing; wireless channel modeling and its applications; and application of machine learning for wireless communications and signal processing.

He was a recipient of 3 Best Paper Awards from IEEE Green Energy and Smart Systems Conference (IGESSC), 2018, 2019 and 2020.
\end{IEEEbiography}

\begin{IEEEbiography}[{\includegraphics[width=1in,height=1.25in,clip]{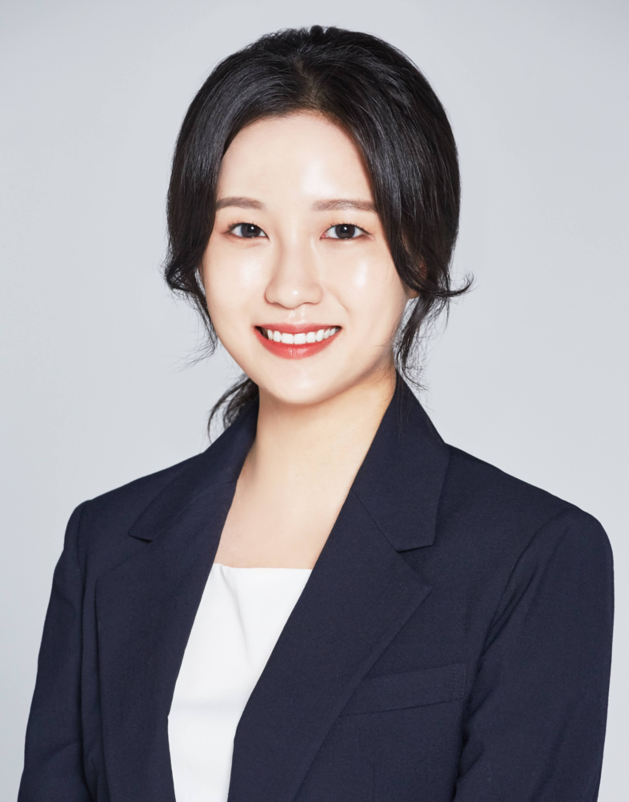}}]{Soyi Jung}
has been an assistant professor at the School of Software, Hallym University, Chuncheon, Republic of Korea, since September 2021. She also holds a visiting scholar position at Donald Bren School of Information and Computer Sciences, University of California, Irvine, CA, USA, from 2021 to 2022. She was a research professor at Korea University, Seoul, Republic of Korea, during 2021. She was also a researcher at Korea Testing and Research (KTR) Institute, Gwacheon, Republic of Korea, from 2015 to 2016. 
She received her B.S., M.S., and Ph.D. degrees in electrical and computer engineering from Ajou University, Suwon, Republic of Korea, in 2013, 2015, and 2021, respectively. 

Her current research interests include network optimization for autonomous vehicles communications, distributed system analysis, big-data processing platforms, and probabilistic access analysis. She was a recipient of Best Paper Award by KICS (2015), Young Women Researcher Award by WISET and KICS (2015), Bronze Paper Award from IEEE Seoul Section Student Paper Contest (2018), ICT Paper Contest Award by Electronic Times (2019), and IEEE ICOIN Best Paper Award (2021).
\end{IEEEbiography}

\begin{IEEEbiography}[{\includegraphics[width=1in,height=1.25in,clip]{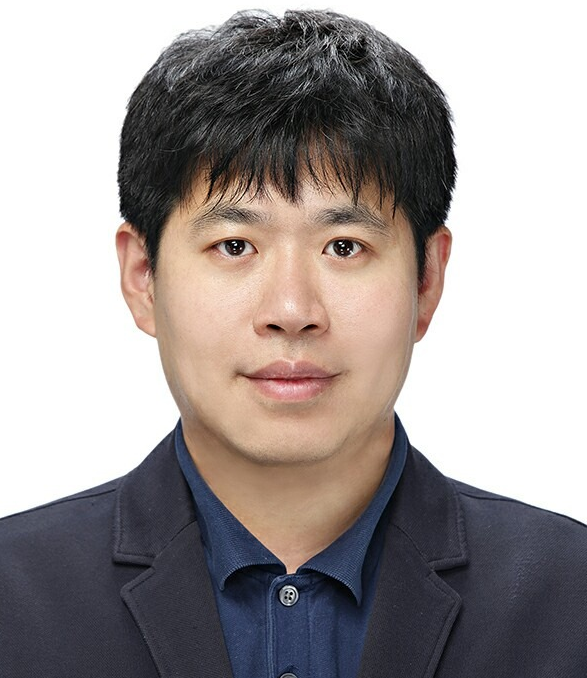}}]{Joongheon Kim}
(M'06--SM'18) has been with Korea University, Seoul, Korea, since September 2019, and he is currently an associate professor at the School of Electrical Engineering. He is also a vice director of the Artificial Intelligence Engineering Research Center at Korea University, Seoul, Korea. 
He received the B.S. and M.S. degrees in computer science and engineering from Korea University, Seoul, Korea, in 2004 and 2006, respectively; and the Ph.D. degree in computer science from the University of Southern California (USC), Los Angeles, California, USA, in 2014. 
Before joining Korea University, he was with LG Electronics CTO Office, Seoul, Korea, from 2006 to 2009; InterDigital, San Diego, California, USA, in 2012; Intel Corporation, Santa Clara in Silicon Valley, California, USA, from 2013 to 2016; and Chung-Ang University, Seoul, Korea, from 2016 to 2019. 

He is a senior member of the IEEE and serves as an associate/guest editor for \textit{IEEE Transactions on Vehicular Technology}, \textit{IEEE Communications Standards Magazine}, and \textit{Computer Networks (Elsevier)}. He is also a distinguished lecturer for \textit{IEEE Communications Society} (2022--2023).

He was a recipient of the Annenberg Graduate Fellowship with his Ph.D. admission from USC (2009), 
Intel Corporation Next Generation and Standards (NGS) Division Recognition Award (2015), Haedong Young Scholar Award by KICS (2018), Paper Awards from IEEE Seoul Section Student Paper Contests (2019 and 2020), \textit{IEEE Systems Journal} Best Paper Award (2020), IEEE ICOIN Best Paper Award (2021), Haedong Paper Award by KICS (2021), and IEEE Vehicular Technology Society (VTS) Seoul Chapter Awards (2019 and 2021).
\end{IEEEbiography}

\end{document}